\documentclass[%
article,
superscriptaddress,
 amsmath,amssymb,
 aps, 
]{revtex4-2}
\usepackage[colorlinks=true,linkcolor=blue]{hyperref}

\usepackage[dvipdfmx]{graphicx}
\usepackage{bm}
\usepackage{amsmath}
\usepackage{amssymb}
\usepackage{color}

\usepackage{subfigure}
\usepackage{xspace}
\usepackage{graphicx}
\usepackage{float}
\usepackage{amsmath}
\usepackage{float}
\usepackage{tikz}
\usepackage{tikz-feynman} 
\usetikzlibrary{shapes,automata,arrows,positioning,calc}
\usetikzlibrary{arrows.meta,bending}
\usepackage{amssymb}
\usepackage{comment}
\usepackage{xcolor}

\newcounter{daggerfootnote}

\newcommand{\ave}[1]{\left\langle #1 \right\rangle}

\newcommand{\imag}{\textrm{i} }

\newcommand{\plaind}{\mathrm{d}}
\newcommand{\dint}[1]{\mathchoice{\!\plaind#1\,}{\!\plaind#1\,}{\!\plaind#1\,}{\!\plaind#1\,}}

\newcommand{\ddintx}[2]{\mathchoice{\!\plaind^{#2}#1\,}{\!\plaind^{#2}#1\,}{\!\plaind^{#2}#1\,}{\!\plaind^{#2}#1\,}}
\newcommand{\dTWOint}[1]{\ddintx{#1}{2}}

\newcommand{\dbar}{\plaind\mkern-6mu\mathchar'26}

\newcommand{\dTWOintbar}[1]{\mathchoice{\!\dbar^{2}#1\,}{\!\dbar^{2}#1\,}{\!\dbar^{2}#1\,}{\!\dbar^{2}#1\,}}

\usepackage{dsfont}

\newcommand{\canetset}[1]{{\mathchoice {\hbox{$\sf\textstyle #1\kern-0.4em #1$}}
{\hbox{$\sf\textstyle #1\kern-0.4em #1$}}
{\hbox{$\sf\scriptstyle #1\kern-0.3em #1$}}
{\hbox{$\sf\scriptscriptstyle #1\kern-0.2em #1$}}}}

\def\nbZ{{\mathchoice {\hbox{$\sf\textstyle Z\kern-0.4em Z$}}
{\hbox{$\sf\textstyle Z\kern-0.4em Z$}}
{\hbox{$\sf\scriptstyle Z\kern-0.3em Z$}}
{\hbox{$\sf\scriptscriptstyle Z\kern-0.2em Z$}}}}

\newcommand{\gpvec}[1]{\textit{\textbf{#1}}}

\newcommand{\zerovec}{\mathbf{0}}
\newcommand{\nullvec}{\zerovec}

\newcommand{\kvec}{\gpvec{k}}

\newcommand{\rvec}{\gpvec{r}}

\newcommand{\uvec}{\gpvec{u}}

\usepackage{bm}
\newcommand{\xivec}{\bm{\xi}}

\newcommand{\ident}{\mathds{1}}

\newcommand{\ptilde}{\tilde{p}}
\newcommand{\ftilde}{\tilde{f}}

\renewcommand{\exp}[1]{\mathchoice{\mathrm{e}^{#1}}{\operatorname{exp}\left(#1\right)}{\operatorname{exp}\left(#1\right)}{\operatorname{exp}\left(#1\right)}}

\newlength \standardfigwidth
\newcounter{exercise}
{\addtocounter{exercise}{1}\begin{center}\begin{minipage}{0.8\linewidth}\textbf{Exercise
\arabic{exercise}:}\begin{itshape}}
{\end{itshape}\end{minipage}\end{center}}

\makeatletter
\newcommand{\creat}[3][]{\@ifempty{#1}{#2^{\dagger}}{\left(#2^{\dagger}\right)^{#1}}\@ifempty{#3}{}{\!(#3)}}

\newcommand{\creatDoi}[3][]{\@ifempty{#1}{\tilde{#2}}{\left(\tilde{#2}\right)^{#1}}\@ifempty{#3}{}{(#3)}}

\newcommand{\annih}[3][]{#2\@ifempty{#1}{}{^{#1}}\@ifempty{#3}{}{(#3)}}

\makeatother

% ...

\newlength{\bibmarkkeyAleft}

\newlength{\bibmarkkeyBleft}

\newlength{\bibmarkkeyCleft}

\newlength{\bibmarkkeyDleft}

\newcommand{\Hermite}[2]{H\! e_{#1}\left(#2\right)}

\begin{document}
\title{Fokker-Planck description of an active Brownian particle with rotational inertia}

\author{Lingyi Wang}
\affiliation{Department of Physics, Wenzhou University, Wenzhou, Zhejiang 325035, China}
\affiliation{Wenzhou Institute, University of Chinese Academy of Sciences, Wenzhou, Zhejiang 325001, China}

\author{Ziluo Zhang}
\affiliation{Department of Physics, Xiamen University, Xiamen, Fujian 361005, China}

\author{Zhongqiang Xiong}
\affiliation{Wenzhou Institute, University of Chinese Academy of Sciences, Wenzhou, Zhejiang 325001, China}

\author{Zhanglin Hou}
\affiliation{Wenzhou Institute, University of Chinese Academy of Sciences, Wenzhou, Zhejiang 325001, China}

\author{Linli He}
\thanks{Corresponding author: \href{mailto:Helinli email}{linlihe@wzu.edu.cn}}
\affiliation{Department of Physics, Wenzhou University, Wenzhou, Zhejiang 325035, China}

\author{Shigeyuki Komura}
\thanks{Corresponding author: \href{mailto:komura@wiucas.ac.cn}{komura@wiucas.ac.cn}}
\affiliation{Wenzhou Institute, University of Chinese Academy of Sciences, Wenzhou, Zhejiang 325001, China}

\begin{abstract}
We develop a perturbative framework to calculate the mean-squared displacement (MSD) of active Brownian 
particles (ABPs) with a finite moment of inertia. 
Starting from the corresponding Fokker-Planck equation, we employ a Fourier transform for the spatial coordinates 
and Hermite polynomials as eigenfunctions for the angular velocity, which enables a systematic perturbative 
expansion of the MSD order by order. 
By resumming the resulting series in Laplace space and performing the inverse transform, we obtain an explicit 
expression for the MSD as a function of the moment of inertia. 
The analytical results are further validated by comparison with numerical simulations.
\\

\end{abstract}

\maketitle

\section{Introduction}

Active matter displays a wealth of non-equilibrium phenomena that are inaccessible in passive systems. 
Among minimal models, active Brownian particles (ABPs) provide a paradigmatic framework for describing 
the interplay between persistent self-propulsion and 
stochastic fluctuations~\cite{Romanczuk12,Stenhammar14,Kurzthaler18}. 
In the standard formulation, ABPs are treated within the overdamped limit, where inertia is neglected due to 
the typically small Reynolds numbers at the micron scale. 
This simplification has nevertheless proven successful in reproducing a wide range of both qualitative 
and quantitative features of active dynamics.

Recent studies have identified regimes in which inertial effects can no longer be 
neglected~\cite{Scholz,Lowen20,Sprenger23,Martins22,Lisin22,Caprini22,Sprenger23,Patel23,Lisin25,Arredondo24}.
This is particularly relevant for granular active matter, mesoscopic or colloidal swimmers in low-viscosity solvents, optically or magnetically driven particles with rapid forcing cycles, and larger synthetic or robotic systems. In these contexts, a finite moment of inertia introduces a temporal memory into the particle's orientational dynamics, directly modifying its rotational persistence. Consequently, inertia qualitatively alters the intermediate-time dynamics, affecting correlations, relaxation processes, and transient states. This emerging recognition motivates a systematic and rigorous extension of the canonical ABP framework to explicitly incorporate rotational inertial effects, forming the foundation for the study of inertial active Brownian particles.

In this work, we address this gap by developing a  analytical theory for ABPs with finite rotational inertia. Our approach extends the underdamped Langevin description to include an inertial term for the angular dynamics, governed by a finite moment of inertia. From this dynamical equation, we derive the corresponding Fokker-Planck equation for the full joint probability distribution function of position, orientation, and angular velocity ~\cite{RiskenBook,Sevilla15}, thereby providing a complete statistical mechanical description of the system.
Analytical expressions for the mean-squared displacement (MSD) are obtained, capturing the short-time 
and long-time regimes. 
To calculate the MSD perturbatively, we develop a recurrence relation in Fourier-Hermite space and introduce an iterative method. 
Our theoretical predictions are validated against numerical simulations of the original Langevin dynamics.
Our results quantify the impact of inertia on transport in active matter and clarify the regions where an overdamped 
theory remains accurate and where inertial effects are important.

\section{Model}
\label{sec:model}

We consider a two-dimensional (2D) circular particle of radius $a$ undergoing self-propelled motion. 
The particle's dynamical state is fully characterized by its position vector $\rvec=[x,y]^{\text{T}}$ 
(``$\text{T}$" indicates transpose) and the orientation unit vector
$\uvec(t)=\uvec(\theta(t))$, where $\theta(t)$ represents the instantaneous angle between the particle's self-propulsion 
direction and the $x$-axis.
The orientation vector is given by $\uvec(t)=[\cos\theta(t),\sin\theta(t)]^{\text{T}}$, which satisfies the normalization 
condition $||\uvec||=1$.
The particle moves with a constant self-propulsion speed $U$ along its instantaneous orientation $\mathbf{u}(t)$. 
For the rotational dynamics, we consider the moment of inertia denoted by $I$. 
The rotational motion is opposed by a viscous torque characterized by the rotational friction coefficient $\gamma$.
On the other hand, we do not consider translational inertia, which requires a separate mathematical treatment.

The ABP is subject to thermal fluctuations. 
The translational and rotational noises, denoted by $\boldsymbol{\xi}_{\mathbf{r}}(t)$ and $\xi_{\theta}(t)$, respectively, 
are Gaussian with zero mean.
We assume that their correlation functions are given by
\begin{subequations}
    \begin{align}
        \ave{\xivec_\rvec(t)}&=\nullvec\ ,\\
        \ave{\xi_{\theta}(t)}&=0 \ ,\\ \ave{\xi_\rvec(t)\xi^{\text{T}}_\rvec(t')}&=2D_{\text{t}}\ident_2\delta(t-t')\ ,\\
        \ave{\xi_{\theta}(t)\xi_{\theta}(t')}&=2D_{\text{r}}\gamma^2\delta(t-t')\ ,
    \end{align}
\end{subequations} 
where $\ident_2$ is a 2D unit matrix, $D_{\rm t}$ is the translational diffusion coefficient,
and $D_{\rm r}$ is the rotational diffusion coefficient.
The latter is related to $\gamma$ through the Einstein relation $D_{\text{r}}=k_{\rm B}T/\gamma$, where 
$k_{\mathrm{B}}$ is Boltzmann constant and $T$ denotes the temperature~\cite{DoiBook}.
In addition, the cross-correlations between $\xivec_\rvec$ and $\xi_{\theta}$ are assumed to be zero.

Under these conditions, the dynamics of the ABP is expressed in terms of a pair of Langevin equations: 
\begin{subequations}
	\begin{align}
		\label{self-P}
	\frac{{\rm d}\rvec}{{\rm d}t}&=U\uvec(t)+\xivec_\rvec(t) \ ,\\
	\label{second order}
	I\frac{{\rm d}^2\theta}{{\rm d}t^2}&=-\gamma\frac{{\rm d}\theta}{{\rm d}t}+\xi_{\theta}(t) \ .
\end{align}
\end{subequations}
By introducing the angular velocity $\omega(t) = {\rm d}\theta/{\rm d}t$, we can rewrite Eq.~(\ref{second order}) in the form
 \begin{equation}
 \label{Langevin omega}
     I\frac{{\rm d} \omega}{{\rm d}t}=-\gamma\omega(t)+\xi_{\theta}(t) \ .
 \end{equation}
In the above model, the noise $\xi_{\theta}(t)$ affects the orientation $\theta(t)$ indirectly through its coupling to the angular velocity $\omega(t)$, rather than acting directly on $\theta(t)$. 
This coupling between the stochastic variables gives rise to nontrivial dynamical behaviors of ABPs~\cite{Romanczuk12}.

\section{Fokker-Planck equation}
\label{sec:FKeq}

A complete statistical description of the particle is provided by the joint probability density function 
$P(\mathbf{r},\theta,\omega;t)$, which denotes the probability density of finding a particle at position $\mathbf{r}$, 
with self-propulsion direction $\theta$ and angular velocity $\omega$ at time $t$, subject to prescribed initial conditions.
To discuss the dynamics of the probability density function, we transform the Langevin equations in Eqs.~(\ref{self-P}) and~(\ref{Langevin omega}) 
into a corresponding Fokker-Planck equation~\cite{RiskenBook}:
\begin{subequations}
\begin{align}
\label{0}
	&\frac{\partial}{\partial t}P(\rvec,\theta,\omega;t)=\mathcal{L}P(\rvec,\theta,\omega;t) \ ,\\
	&\mathcal{L}=-U\uvec(t)\cdot\nabla_{\rvec}+D_{\text{t}}\nabla_{\rvec}^2-\omega\frac{\partial}{\partial\theta}+\frac{\gamma}{I}\frac{\partial}{\partial\omega}\omega+\frac{D_{\text{r}}\gamma^2}{I^2}\frac{\partial^2}{\partial\omega^2} \ ,
\label{1}	
\end{align}
\end{subequations}
where $\nabla_{\rvec}$ in the operator $\mathcal{L}$ denotes the 2D spatial gradient operator.

We next determine the eigenfunctions associated with the last two terms on the right-hand side of Eq.~(\ref{1}). 
They are given by
\begin{equation}
	\label{3}
	\frac{\gamma}{I}\frac{\partial}{\partial\omega}\left[\omega f_n(\omega)\right]+\frac{\gamma^2 D_{\text{r}}}{I^2}\frac{\partial^2}{\partial\omega^2}f_n(\omega)=-\frac{\gamma}{I}n f_n(\omega) \ ,
\end{equation}
where
\begin{subequations}
\begin{align}
\label{Hermite}
	f_n(\omega)&=\exp{-\omega^2/2L^2}H\!e_{n}\left(\frac{\omega}{L} \right)\ ,\\
    \ftilde_n(\omega)&=\frac{1}{\sqrt{2\pi}n!}H\!e_{n}\left(\frac{\omega}{L}\right)\ .
\end{align}
\end{subequations}
Here, we have introduced $L=\sqrt{\gamma D_{\text{r}} /I}$, and 
$H\!e_{n}(x)$ is the $n$-th order of the 
\textit{probabilist's Hermite polynomials} ($n\in Z^{+}_0$)~\cite{AbramowitzStegun,Garcia-Millan21,Zhang24}.
Our definition of $\Hermite{n}{x}$ corresponds to $2^{-n/2}$\texttt{HermiteH[$n,x/\sqrt{2}$]}~\cite{Mathematica}. 
In the above, $f$ and $\tilde{f}$ also satisfy an orthogonality relation
\begin{equation}
    \int \dint{\omega} \, f_n(\omega) \ftilde_m(\omega)=L\delta_{n,m}\ ,
\end{equation}
where $\delta_{n,m}$ is the Kronecker delta.

The following relations of the Hermite polynomials will be employed in the subsequent analysis:
\begin{subequations}
\label{eq:Hermite_properties}
\begin{align}
	&\int_{-\infty}^{\infty}\dint \omega \exp{-\omega^2/2L^2}H\!e_{n} \left(\frac{\omega}{L}\right)H\!e_{m}\left(\frac{\omega}{L}\right)=L\sqrt{2\pi}n!\delta_{n,m} \ ,\\
	&\frac{\omega}{L}H\!e_{n}\left(\frac{\omega}{L}\right)=H\!e_{n+1}\left(\frac{\omega}{L}\right)+n H\!e_{n-1}\left(\frac{\omega}{L}\right) \ .
\end{align}
\end{subequations}

\section{Mean-squared displacement}

\subsection{Fourier transformation}
\label{sec:fourier}

The main objective of this work is to calculate the MSD of a 2D ABP in the presence of the moment of inertia $I$. 
The MSD is defined as
\begin{equation}
\label{msd}
    \ave{(\rvec-\rvec_0)^2(t-t_0)}=\int \dTWOint{r} \,\dTWOint{r_0} \int \dint{\omega} \, \dint{\omega_0} \int_0^{2\pi} 
    \dint{\theta} \, \dint{\theta_0} (\rvec-\rvec_0)^2 P(\rvec,\theta,\omega;t\big|\rvec_0,\theta_0,\omega_0;t_0) 
    P(\rvec_0,\theta_0, \omega_0)\ ,
\end{equation}
where $\rvec_0$, $\theta_0$, and $\omega_0$ denote the initial values of the position, orientation, and angular velocity, 
respectively, at the initial time $t_{0}$, 
$P(\rvec,\theta,\omega;t\big|\rvec_0,\theta_0,\omega_0;t_0)$ is the conditioned probability density function,
and $P(\rvec_0,\theta_0, \omega_0)$ is the initial probability density function at $t=t_{0}$.

To circumvent the explicit evaluation of the integral, we transform the probability density function as
\begin{align}
\label{p expansion}
P(\rvec,\theta,\omega;t\big|\rvec_0,\theta_0,\omega_0;t_0) & = \frac{1}{2\pi L} \sum_{\alpha,\alpha'=-\infty}^\infty \sum_{n,n'=0}^\infty \int \dTWOintbar{k}\, \exp{\imag [\kvec\cdot (\rvec-\rvec_0)]}\exp{\imag (\alpha \theta-\alpha'\theta_0)}   
\nonumber \\
& \times f_n(\omega) 
\ftilde_{n'}(\omega_0) \tilde{p}_{\alpha,n,\alpha',n'}(\kvec;t)\exp{-(D_{\rm t} k^2+\gamma n/I)(t-t_0)}  \ ,
\end{align}
where $\dbar=\dint/(2\pi)$ and $\kvec=[k_x,k_y]^{\text{T}}$ is the 2D wave-vector. 
Throughout this paper, we use Roman letters to denote angular velocity indices, while Greek letters are employed for the orientation indices.

By substituting Eq.~(\ref{p expansion}) into Eq.~(\ref{msd}), we first perform the integrals over the initial conditions. 
A suitable choice of the initial distribution is made to simplify the subsequent calculations.
We assume a delta function for the position, $P(\mathbf{r}_{0}) = \delta(\mathbf{r}_{0})$,
a uniform distribution for the orientation, $P(\theta_{0}) = 1/(2\pi)$,
and a Gaussian distribution for the angular velocity, 
$P(\omega_0)=(\sqrt{2\pi}L)^{-1} \exp{-\omega_0^2/2L^2}$.
Then, we have  
\begin{subequations}
    \begin{align}
        P(\rvec,\theta,\omega;t)&= \int \dTWOint{r_0} \int \dint{\omega_0} \int_0^{2\pi}\dint{\theta_0}\, P(\rvec,\theta,\omega;t\big|\rvec_0,\theta_0,\omega_0;t_0) P(\rvec_0,\theta_0, \omega_0)
\\&=\frac{1}{2\pi L} \sum_{\alpha,\alpha'=-\infty}^\infty \sum_{n,n'=0}^\infty \int \dTWOint{r_0} \int \dint{\omega_0} \int_0^{2\pi}\dint{\theta_0}\int \dTWOintbar{k}\, \exp{\imag [\kvec\cdot (\rvec-\rvec_0)]}\exp{\imag (\alpha \theta-\alpha'\theta_0)}   f_n(\omega) 
\ftilde_{n'}(\omega_0) \tilde{p}_{\alpha,n,\alpha',n'}(\kvec;t) 
\nonumber \\
&\times\delta(\rvec_0)\frac{1}{2\pi} \frac{1}{\sqrt{2\pi}L}\exp{-\omega_0^2/2L^2} \label{initial_condition_integral}\exp{-(D_{\rm t} k^2+\gamma n/I)(t-t_0)}\\
&=\frac{1}{2\pi L} \sum_{\alpha,\alpha'=-\infty}^\infty \sum_{n,n'=0}^\infty \int\dTWOintbar{k} \, \exp{\imag \kvec\cdot \rvec}\exp{\imag \alpha \theta}  f_n(\omega) 
\ftilde_{n'}(\omega_0) \tilde{p}_{\alpha,n,\alpha',n'}(\kvec;t) \delta_{\alpha',0}\delta_{n',0}\exp{-(D_{\rm t} k^2+\gamma n/I)(t-t_0)}  \ ,
\end{align}
\end{subequations}
where the three initial distribution functions of $\rvec_0$, $\theta_{0}$, and $\omega_{0}$ are included in 
Eq.~(\ref{initial_condition_integral}).

We employ the orthogonality properties of the Hermite polynomials and exponential functions, 
and fix the indices corresponding to the initial conditions to zero. 
For notational simplicity, we introduce 
$\tilde{p}_{\alpha,n}=\tilde{p}_{\alpha,n,0,0}(\kvec;t)$ for later calculations.
By performing the remaining integrals, one can evaluate the MSD by an expression~\cite{Sevilla15} 
\begin{align}
	\label{msd1}
	\langle\rvec^2(t)\rangle= 4D_{\text{t}}t-[ \nabla_{\kvec}^2 \tilde{p}_{0,0}(\kvec;t)]_{\kvec=\nullvec} \ ,
\end{align}
where $\nabla_{\kvec}^2$ is the Laplacian in the Fourier space. 
From Eq.~(\ref{msd1}), the MSD can be obtained only from $\tilde{p}_{0,0}$ by taking the second derivative with 
respect to  $\kvec$ and subsequently setting $\mathbf{k}=\nullvec$. 
In the following, we demonstrate the procedure for calculating $\tilde{p}_{0,0}$.

\subsection{Iteration and perturbation method}
\label{sec:iteration}

We first derive a recurrence relation for $\tilde{p}_{\alpha,n}$ in order to calculate $\tilde{p}_{0,0}$. 
By substituting the eigenfunction expansion in Eq.~(\ref{p expansion}) into the Fokker-Planck equation
in Eq.~(\ref{0}), and by employing the properties of the Hermite polynomials in Eq.~(\ref{eq:Hermite_properties}), we obtain
\begin{align}
	\label{8}
	\frac{{\rm d}}{{\rm d}t}\tilde{p}_{\alpha,n}&=-\frac{\imag U k}{2}\left(\tilde{p}_{\alpha-1,n}\exp{-\imag \psi}+\tilde{p}_{\alpha+1,n}\exp{\imag \psi}\right)-\imag \alpha L\left[\tilde{p}_{\alpha,n-1}\exp{\gamma t/I}+(n+1)\tilde{p}_{\alpha,n+1}\exp{-\gamma t/I}\right] \ ,
\end{align}
where we have introduced the wave-vector decomposition $k_x \pm \imag k_y = k\exp{\pm \imag \psi}$
with $k=||\kvec||$.
By taking the second time derivative of Eq.~(\ref{msd1}) and substituting all occurrences of  ${\rm d}\ptilde/{\rm d}t $ 
using Eq.~(\ref{8}), we obtain
\begin{align}
	\frac{{\rm d}^2}{{\rm d}t^2}\langle\rvec^2(t)\rangle=&\frac{U^2}{2}
	[\nabla_{\kvec}^2 (k_x^2+k_y^2)\tilde{p}_{0,0}]_{\kvec=\nullvec}+\frac{U^2}{4}\exp{-4D_{\text{r}}t}\{\nabla_{\kvec}^2[(k_x+\imag k_y)^2\tilde{p}_{2,0}+(k_x-\imag k_y)^2\tilde{p}_{-2,0}]\}_{\kvec=\nullvec}\nonumber\\
	-&[\nabla^2_{\kvec}\frac{UkL}{2}\exp{-\gamma t/I}(\tilde{p}_{-1,1}\exp{-\imag\psi}-\tilde{p}_{1,1}\exp{\imag\psi})]_{\kvec=\nullvec} \ .
\label{secondMSD}	
\end{align}
Since $\ptilde_{0,0}(\nullvec)=1$ as a consequence of the normalization of the probability density function, the first term 
on the right-hand side reduces to $2U^{2}$.
Because $\nabla^2_{\kvec} (k_x\pm\imag k_y)^2=0$, the second term in the right-hand side of Eq.~(\ref{secondMSD})
vanishes after we set $\kvec=\nullvec$.
Hence, we obtain
\begin{align}
\label{Eq:zero-th order}
	\frac{{\rm d}^2}{{\rm d}t^2}\langle\rvec^2(t)\rangle=2U^2 -
	[\nabla^2_{\kvec}\frac{UkL}{2}\exp{-\gamma t/I}(\tilde{p}_{-1,1}\exp{-\imag\psi}-\tilde{p}_{1,1}\exp{\imag\psi})]_{\kvec=\nullvec} \ .
\end{align}

Due to the presence of the mode $\ptilde_{\pm1,1}$, the MSD cannot be obtained in a straightforward manner. 
To circumvent this difficulty, we further take the time derivatives of Eq.~(\ref{Eq:zero-th order}), 
expand the term ${\rm d}\tilde{p}/{\rm d}t$ using Eq.~(\ref{8}), and neglect all higher-order contributions, 
namely those with $n>2$ and $|\alpha|>1$. 
The first three contributions are listed below:
\begin{subequations}
\begin{align}
	\label{12}
	&\frac{{\rm d}^2}{{\rm d}t^2}\langle\rvec^2(t)\rangle_0=2U^2-
	[\nabla^2_{\kvec}\frac{UkL}{2}\exp{-\gamma t/I}(\tilde{p}_{-1,1}\exp{-\imag\psi}-\tilde{p}_{1,1}\exp{\imag\psi})]_{\kvec=\nullvec} \ , \\
	\label{13}
	&\frac{{\rm d}^4}{{\rm d}t^4}\langle\rvec^2(t)\rangle_1+\frac{\gamma}{I}\frac{{\rm d}^3}{{\rm d}t^3}\langle\rvec^2(t)\rangle_1=2U^2L^2\imag^2 -
	[\nabla^2_{\kvec}\frac{\imag^2 UkL^3}{2}\exp{-\gamma t/I}(\tilde{p}_{-1,1}\exp{-\imag\psi}-\tilde{p}_{1,1}\exp{\imag\psi})]_{\kvec=\nullvec} \ , \\
	\label{14}
	&\frac{{\rm d}^6}{{\rm d}t^6}\langle\rvec^2(t)\rangle_2+\frac{2\gamma}{I}\frac{{\rm d}^5}{{\rm d}t^5}\langle\rvec^2(t)\rangle_2+\frac{\gamma^2}{I^2}\frac{{\rm d}^4}{{\rm d}t^4}\langle\rvec^2(t)\rangle_2=2U^2L^4\imag^4 -
	[\nabla^2_{\kvec}\frac{\imag^4 UkL^5}{2}\exp{-\gamma t/I}(\tilde{p}_{-1,1}\exp{-\imag\psi}-\tilde{p}_{1,1}\exp{\imag\psi})]_{\kvec=\nullvec} \ ,
\end{align}
\end{subequations}
where the subscript $n$ of $\langle \rvec^2 \rangle_n$ indicates the number of iterations performed.
The result for the $n$-th iteration is
\begin{subequations}
\begin{align}
    \frac{{\rm d}^{n+2}}{{\rm d}t^{n+2}}\sum_{k=0}^{n}\binom{n}{k}\left(\frac{{\rm d}}{{\rm d}t}\right)^{n-k}\left(\frac{\gamma}{I}\right)^k \langle \rvec^2(t)\rangle_n
    &=\frac{{\rm d}^{n+2}}{{\rm d}t^{n+2}}\left(\frac{{\rm d}}{{\rm d}t} + \frac{\gamma}{I}\right)^n \langle \rvec^2(t)\rangle_n\\
    &=2U^2L^{2n}\imag^{2n}-
    [\nabla^2_{\kvec}\frac{\imag^{2n}UkL^{2n+1}}{2}\exp{-\gamma t/I}(\tilde{p}_{-1,1}\exp{-\imag\psi}-\tilde{p}_{1,1}\exp{\imag\psi})]_{\kvec=\nullvec} \ ,
\label{eq17b}
    \end{align}
\end{subequations}
where $\binom{n}{k}=n!/[k!(n-k)!]$.
Hence, the MSD can be calculated as the summation of $\ave{\rvec^2}_n$ over $n$ and becomes 
\begin{equation}
\langle\rvec^2(t)\rangle=4D_{\text{t}}t+\sum_{n=0}^\infty \langle\rvec^2(t)\rangle_n \ .
\end{equation}

Next, we apply the Laplace transform with respect to time $t$, adopting the following convention:
\begin{equation}
y_n(s)=\int_{0}^{\infty}{\rm d}t\, \langle \rvec^2(t)\rangle_n\exp{-st} \ ,
\end{equation}
and thereby transform the differential equations into a set of linear equations given by 
\begin{equation}
	s^{n+2}\left(s+\frac{\gamma}{I}\right)^n y_n(s)=\frac{2U^2 L^{2n}\imag^{2n}}{s} \ .
\end{equation}

Since the last term in Eq.~(\ref{eq17b}) contributes to the next iteration, we only sum the first term in Eq.~(\ref{eq17b}).
By summing all orders, we obtain the Laplace transform of the MSD $y(s)$ as
\begin{align}
	y(s)&=\sum_{n=0}^\infty y_n(s)=\sum_{n=0}^{\infty}\frac{2U^2}{s^3}\left[\frac{-L^2}{s(s+\gamma/I)}\right]^n
    =\frac{2U^2}{s^3}\frac{s(s+\gamma/I)}{s(s+\gamma/I)+L^2} \ ,
\end{align}
where the summation is evaluated using geometric series.
Transforming $y(s)$ back to the time domain, we obtain the MSD as
\begin{align}
\langle\rvec^2(t)\rangle &=4D_{\rm t}t+\frac{2IU^2}{\gamma D_{\rm r}}\left(1-\frac{\gamma}{ID_{\rm r}}\right) +\frac{2U^2 t}{D_{\rm r}}
\nonumber \\
& +\frac{2U^2(\gamma^2-I\gamma D_{\rm r})}{\gamma^2D_{\rm r}^2}\exp{-\gamma t/2I} \cosh(\Gamma t)
+\frac{U^2(2\gamma^2/I-6\gamma D_{\rm r})}{2\Gamma\gamma D_{\rm r}^2}\exp{-\gamma t/2I}\sinh(\Gamma t)+\mathcal{O}\left(\frac{ID_{\rm r}}{\gamma}\right)\ , 
\label{fullMSD}
\end{align}
where $\Gamma=\sqrt{(\gamma/2I)^2-\gamma D_{\text{r}}/I}$. 
When $\Gamma$ is purely imaginary, the hyperbolic functions can be expressed in terms of trigonometric functions, 
i.e., $\cosh{(\Gamma t)}=\cos{(\Im(\Gamma)t)}$ and $\sinh{(\Gamma t)}/\Gamma=\sin{(\Im(\Gamma)t)}/\Im(\Gamma)$. 
Since higher-order contributions of $\tilde{p}_{\alpha,n}$ have been neglected, the above perturbative result is exact 
only up to order $\sqrt{I D_{r}/\gamma}$. 
Therefore, we have included a correction term of order $\mathcal{O}(ID_{\rm r}/\gamma)$ in Eq.~(\ref{fullMSD}).

\subsection{Asymptotic limits}
\label{sec:analysis}

For convenience, we introduce the dimensionless parameter $\beta=\sqrt{ID_{\rm r}/\gamma}$.
Then, the obtained MSD in Eq.~(\ref{fullMSD}) can be rewritten as 
\begin{align}
\ave{\rvec^2(t)} & =4D_{\rm t}t+\frac{2U^2}{D_{\rm r}^2}(-1+\beta)+\frac{2U^2 t}{D_{\rm r}}
\nonumber \\
& +(1-\beta)\frac{2U^2}{D_{\rm r}^2}e^{-D_{\rm r}t/2\beta}\cosh \left(
{\frac{D_{\rm r}\sqrt{1-4\beta}}{2\beta}t} \right)
+\frac{1-3\beta}{\sqrt{1-4\beta}}\frac{2U^2}{D_{\rm r}^2}e^{-D_{\rm r}t/2\beta}
\sinh \left( {\frac{D_{\rm r}\sqrt{1-4\beta}}{2\beta} t}\right) +\mathcal{O}(\beta^2) \ .
\label{finalMSD}
\end{align}
In the short-time limit $t \to 0$, using the Taylor expansion of the time-dependent terms, we recover diffusive and 
ballistic behaviors from the leading-order contribution as 
\begin{equation}
\langle \rvec^2(t)\rangle=4D_{\text{t}} t +U^2 t^2 +\mathcal{O}(t^3) \ .
\label{short-time}    
\end{equation}
In the long-time limit $t \to \infty$, on the other hand, the last two terms in Eq.~(\ref{finalMSD}) vanish due to exponential decay. 
Hence, the effective diffusion coefficient can be obtained as 
\begin{equation}
D_{\text{eff}}=\lim_{t\rightarrow \infty}\frac{\langle \rvec^2(t) \rangle}{4t}=D_{\text{t}}+\frac{U^2}{2D_{\rm r}}+\mathcal{O}(\beta^2) \ .
\label{long-time}
\end{equation}
Note that both Eqs.~(\ref{short-time}) and~(\ref{long-time}) are independent of $\beta$ in the short-time 
and long-time limits, respectively.

If the moment of inertia is sufficiently small, $\beta \ll 1$, we can further expand Eq.~(\ref{finalMSD}) up to the leading order 
in $\beta$ as 
\begin{align}
\label{first_order_correction}
\langle\rvec^2(t)\rangle=4D_{\text{t}}t+\frac{2U^2}{D_\text{r}^2}\left(\exp{-D_{\text{r}}t}-1+D_{\text{r} }t\right)+\beta \frac{2U^2}{D_\text{r}^2}\left[ 1-(1+D_\text{r} t)\exp{-D_{\text{r}}t}\right] +\mathcal{O}(\beta^2)\ ,
\end{align}
When $\beta \rightarrow 0$ (or $I = 0$), the MSD reduces to 
\begin{align}
\lim_{\beta\rightarrow 0}\langle \rvec^2(t)\rangle &=4D_{\text{t}}t+\frac{2U^2}{D_\text{r}^2}\left(\exp{-D_{\text{r}}t}-1+D_{\text{r} }t\right) \ .
\end{align}
which coincides with the result for standard ABPs, as it should~\cite{Romanczuk12,Lowen20}.
Since the last term in Eq.~(\ref{first_order_correction}) is non-negative, rotational inertia always increases 
the MSD compared to the case without inertia.
Physically, this corresponds to particles orienting slowly due to their moment of inertia.
However, taking the short-time limit $t\rightarrow 0$ in Eq.~(\ref{first_order_correction}) does not reproduce the diffusive 
and ballistic behaviors in Eq.~(\ref{short-time}).
Because of the exponential and hyperbolic terms in Eq.~(\ref{finalMSD}), the limit $\beta\rightarrow 0$ is singular, and 
the resulting asymptotic expansion is non-uniform, namely, the two limits $\beta\rightarrow0$ and $t\rightarrow 0$ do 
not commute in this case.  
Consequently, Eq.~(\ref{first_order_correction}) is valid only for sufficiently large $t$, and the short-time behavior cannot be 
obtained from it.

\begin{figure}[htb]
\centering
\includegraphics[width=0.9\textwidth]{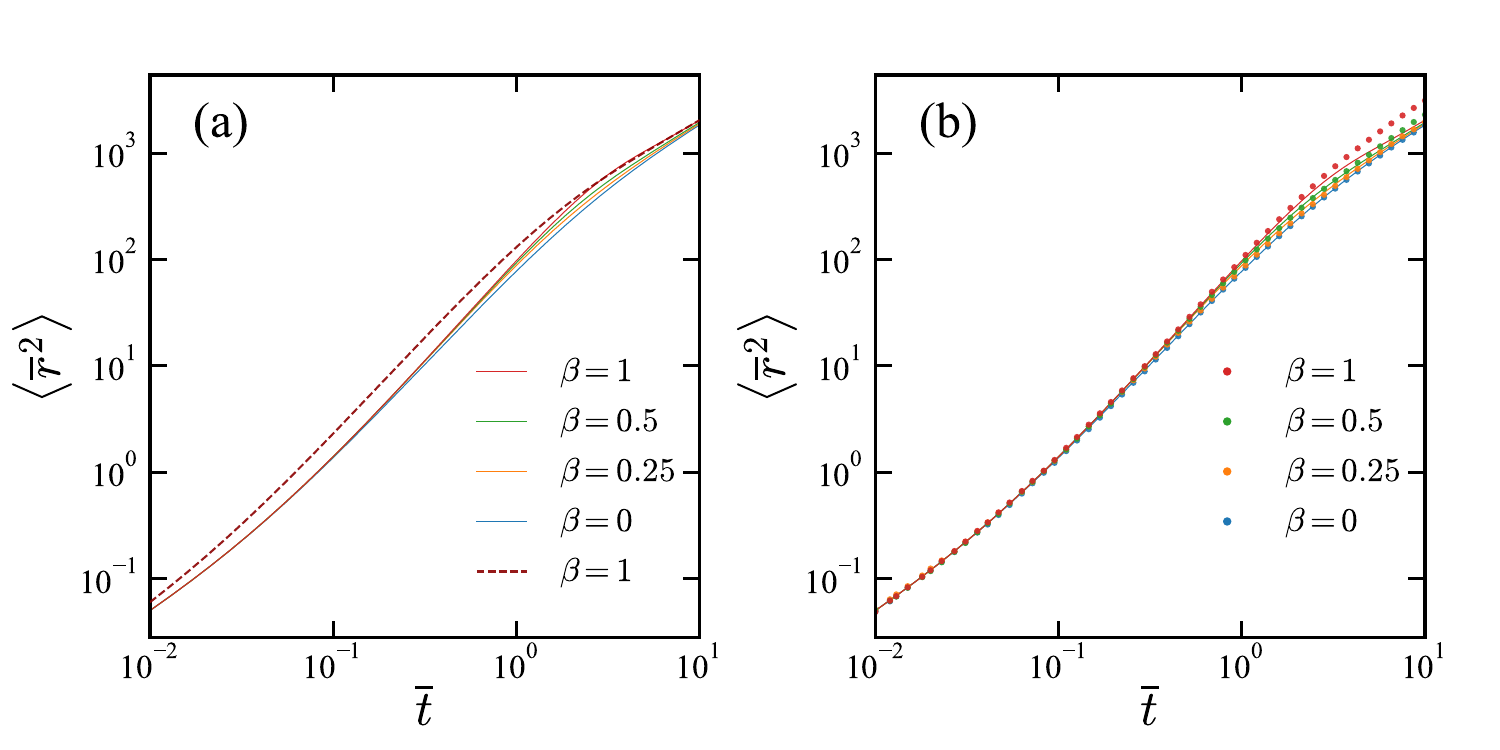}
\caption{(a) Dimensionless MSD $\langle\overline{\rvec}^2\rangle=\langle \rvec^2\rangle D_{\rm r}/D_{\rm t}$ as 
a function of dimensionless time $\overline{t}=D_{\rm r}t$ when the dimensionless moment of inertia is 
changed as $\beta=\sqrt{ID_{\rm r}/\gamma}= 0, 0.25, 0.5, 1$.
The dimensionless self-propulsion velocity is chosen as $\overline{U}=U/\sqrt{D_{\rm t} D_{\rm r}}=10$.
The solid lines correspond to Eq.~(\ref{finalMSD}), and the dashed line corresponds to Eq.~(\ref{first_order_correction})
when $\beta=1$.}
(b) Comparison between the analytical and numerical simulation results. 
The parameters and the solid lines are identical to those in (a).
The symbols represent simulation data obtained by numerically solving Eqs.~(\ref{self-P}) and~(\ref{second order}) 
using the Euler-Maruyama method.

\label{MSDplot}	
\end{figure}

In Fig.~\ref{MSDplot}(a), we plot the dimensionless MSD $\langle\overline{\rvec}^2\rangle=\langle \rvec^2\rangle D_{\rm r}/D_{\rm t}$ 
in Eq.~(\ref{finalMSD}) as a function of dimensionless time $\overline{t}=D_{\rm r}t$ for different values of $\beta$. 
Here, the dimensionless self-propulsion velocity is chosen to be $\overline{U}=U/\sqrt{D_{\rm t} D_{\rm r}}=10$.
As $\beta$ increases from zero, the MSD becomes larger in the time regime 
$10^{-1} \leq \overline{t} \leq 10$.
In the short-time and long-time limits, however, the MSD is independent of $\beta$, as can be confirmed by 
Eqs.~(\ref{short-time}) and~(\ref{long-time}).
We also plot the first-order approximation in Eq.~(\ref{first_order_correction}) for $\beta=1$, shown as a 
dashed line for comparison.
As discussed above, the first-order approximation cannot capture the ballistic regime because of the essential singularity 
associated with $e^{-1/\beta}$.

In Fig.~\ref{MSDplot}(b), we present a comparison between the analytical results and numerical simulations. 
In the simulations, we employ the Euler-Maruyama method to numerically solve the Langevin stochastic differential 
equations in Eqs.~(\ref{self-P}) and~(\ref{second order}). 
Position data from 2,000 trajectories are generated to evaluate the MSD statistically. 
Across the entire time range, the simulated data points show good agreement with the theoretical predictions, 
while minor discrepancies appear between the numerical and analytical results as the inertia increases.
This is because our analytical result is valid when the moment of inertia (or $\beta$) is small enough. 
The result is valid only up to first order in $\beta$, i.e., we retain only a finite number of terms in the Hermite  
and Fourier expansions.
Our theory can reproduce the simulation results more accurately by including higher-order terms, at the cost of 
increased computational effort.

\section{Summary and discussion}

We have developed a perturbative framework to obtain approximate solutions for the MSD of a 2D ABP with moment of inertia. 
Starting from the Langevin dynamics, we reformulated the problem in terms of a Fokker-Planck equation, employed Fourier 
transforms for spatial coordinates, and used Hermite polynomials for angular velocity modes. 
This allowed us to derive a recurrence relation and truncate the hierarchy, leading to an approximate expression for the MSD 
in Laplace space. 
The final MSD was obtained by transforming back to the time domain. 
In the limit of vanishing inertia, our formulation reduces to that of the standard overdamped ABP
model~\cite{Romanczuk12,Lowen20}.

We have validated the analytical results against numerical simulations across different timescales. 
For $I=0$, the MSD exhibits the expected linear growth in both short- and long-time limits, with a quadratic regime emerging 
at intermediate times due to self-propulsion.
For finite inertia ($I>0$), particles display ballistic behavior at short times before crossing over to linear diffusion at long times. 
This demonstrates that inertia introduces a distinct intermediate dynamical regime, highlighting a difference between 
inertial and overdamped ABPs.

A natural direction for future work is a unified treatment of translational and rotational inertia. 
In the inertial regime, Hermite polynomials form the eigenfunction of velocity fields~\cite{Zhang24}.
Computing the MSD is then nontrivial because underdamped dynamics generate additional position-velocity cross terms. 
Nevertheless, the long-time effective diffusion coefficient can be obtained via the Green-Kubo relation 
because the velocity is not dependent on the position variables, and the latter can be ignored.

The present framework can be extended in several directions. 
Incorporating interparticle interactions, external confinement, or shear flows would enable the study of collective 
phenomena and transport in more realistic environments. 
Moreover, connecting the theory with experimental systems such as granular active matter, colloids at intermediate Reynolds numbers, or synthetic microswimmers under rapid driving could provide direct tests of inertial effects. 
These extensions would further clarify the role of inertia in active matter.

\begin{acknowledgements}

We thank K.\ Yasuda for helpful discussions.
L.H.\ and S.K.\ acknowledge the support by the National Natural Science Foundation of China 
(Nos.\ 12104453, 22273067, and 12274098).
S.K.\ acknowledges the startup grant of Wenzhou Institute, 
University of Chinese Academy of Sciences (No.\ WIUCASQD2021041). 
S.K.\ acknowledges the support by the Japan Society for the Promotion of Science (JSPS) Core-to-Core 
Program ``Advanced core-to-core network for the physics of self-organizing active matter" (No.\ JPJSCCA20230002).
L.Y.\ and Z.Z.\ contributed equally to this work.
\end{acknowledgements}


\end{document}